# Stepwise quantized surface states and delayed Landau level hybridization in Co cluster-decorated BiSbTeSe$_2$ topological insulator devices

Shuai Zhang[1], Li Pi[2,3], Rui Wang[1], Geliang Yu[1], Xing-Chen Pan[1], Zhongxia Wei[1], Jinglei Zhang[3], Chuanying Xi[3], Zhanbin Bai[1], Fucong Fei[1], Mingyu Wang[1], Jian Liao[4], Yongqing Li[4], Xuefeng Wang[5], Fengqi Song[1,*], Yuheng Zhang[3*], Baigeng Wang[1*], Dingyu Xing[1] and Guanghou Wang[1]

[1]National Laboratory of Solid State Microstructures, College of Physics and Collaborative Innovation Center of Advanced Microstructures, Nanjing University, Nanjing 210093, China

[2]Hefei National Laboratory for Physical Sciences at Microscale, University of Science and Technology of China, Hefei 230026, China

[3]High Magnetic Field Laboratory, Chinese Academy of Sciences and Collaborative Innovation Center of Advanced Microstructures, Hefei, Anhui 230031, China

[4]Institute of Physics, Chinese Academy of Sciences, Beijing, China

[5]National Laboratory of Solid State Microstructures, Collaborative Innovation Center of Advanced Microstructures, and School of Electronic Science and Engineering, Nanjing University, Nanjing, 210093,China

---

[*] Corresponding authors: songfengqi@nju.edu.cn, zhangyh@ustc.edu.cn, bgwang@nju.edu.cn. Fax: +86-25-83595535. The first 3 authors contributed equally.

In three-dimensional topological insulators (TIs), the nontrivial topology in their electronic bands casts a gapless state on their solid surfaces, [1, 2, 3, 4, 5, 6, 7, 8, 9, 10, 11], using which dissipationless TI edge devices based on the quantum anomalous Hall (QAH) effect [12, 13, 14, 15] and quantum Hall (QH) effect [16, 17, 18, 19] have been demonstrated. Practical TI devices present a pair of parallel-transport topological surface states (TSSs) on their top and bottom surfaces. However, due to the no-go theorem[20, 21], the two TSSs always appear as a pair and are expected to quantize synchronously. Quantized transport of a separate Dirac channel is still desirable, but has never been observed in graphene even after intense investigation over a period of 13 years [25-29], with the potential aim of half-QHE[2]. By depositing Co atomic clusters, we achieved stepwise quantization of the top and bottom surfaces in BiSbTeSe$_2$ (BSTS) TI devices. Renormalization group flow diagrams[13, 22] (RGFDs) reveal two sets of converging points (CVPs) in the ($G_{xy}$, $G_{xx}$) space, where the top surface travels along an anomalous quantization trajectory while the bottom surface retains 1/2 $e^2/h$. This results from delayed Landau-level (LL) hybridization (DLLH) due to coupling between Co clusters and TSS Fermions.

**The Quantum Hall (QH) effect of the surface-dominated TI device**

We fabricated field-effect-transistor BSTS[10] devices using a standard lift-off procedure (See Methods). The Co clusters were deposited on the bare devices using a cluster beam source, as described elsewhere[23]. A schematic of our devices is shown in the inset of **Figure 1(a)**. Successful deposition of nanoclusters was shown by elemental energy-dispersive X-ray spectroscopy. Fig. 1(a) also shows the typical temperature ($T$)-dependent resistance of the devices. An insulating behavior can be seen at high temperatures (above 100 K), which is attributed to the transport of the bulk electronic state, and is then followed by a metallic increase at low temperatures. This is similar to the case for pristine BSTS devices as reported by other groups[11, 16]. The maximum bulk resistivity reached more than 10 Ω cm. An atomic force microscopic image (Fig. 1(b)) shows nano-islands with diameters of a few tens of nanometers and a height of around 5 nm. The transport parameters of the devices are listed in the supplementary material.

In Fig. 1(c), the back gate voltage ($V_g$)-dependent longitudinal conductance ($G_{xx}$) was measured at a series of temperatures from 2 to 40 K. All the transfer curves show bipolar transport behaviors, where the minimum conductance corresponds to the Dirac point (DP) of the bottom surface. At $T = 2$ K, the DP is around $V_g = 16$ V and the minimum conductance is about 5 $e^2$/h, where $e$ is the electron charge and $h$ is the Planck constant. Furthermore, $G_{xx}$ exhibits a linear dependence on the gate voltages[17, 24]. We measured a number of samples, some also prior to cluster deposition. We plotted the sheet resistance ($R_{sh}$) for all of them at 2 K (red dots) and 270 K (blue dots) in Fig. 1(d), where we show that all values of $R_{sh}$ at 2 K fall into a small range independent of thickness. This implies surface-dominant transport in our devices, allowing the successful observation of the QH effect.

A high magnetic field drives the device towards the QH state. In Fig. 1(e), for a magnetic field of 12 Tesla (T), the Hall conductance ($G_{xy}$) steadily reaches 0 and -1 $e^2/h$ while the backgate voltage is varied from 0 to 40 V. At the same time, the longitudinal conductances ($G_{xx}$) fluctuate and show local minimums at both the Hall plateaus. Similar QH effects [16, 17, 18, 19] have been interpreted based on the fact that the two TSSs (top and bottom) each contribute a half-integer plateau. The DP for the bottom surface is $V_g$ = 20 V (Table 1). Hence, we assign a half-integer index ν for each surface for both plateaus[16, 17, 19]. $ν_b$ (bottom surface) is set at 1/2 and -1/2 respectively, and $ν_t$ (top surface) is maintained at -1/2. The top surface is p-type.

**The anomalous quantization trajectory of a single Dirac channel extracted by RGFD**

Quantized transport of a single channel has been long desired for Dirac Fermion devices. For graphene, integer QH plateaus described by the series (4n + 2) $e^2/h$ have been observed[25, 26], due to the spin and sublattice degeneracy of the material[27, 28, 29]. QH experiments on TI devices suffer from parallel transport of the two surface states and result in QH plateaus of ($n_T + n_B + 1$) $e^2/h$. However, it is interesting to note that in our cluster-decorated device, the half-integer -3/2 Hall plateau appeared when measuring the device at a medium field of -7 T, as marked by the arrow in **Figure 2(a)**, in which a plateau of zero is also seen. Both plateaus are supported by fixed $G_{xy}$ values and local minimums of $G_{xx}$. Measurements made on other devices show the QH plateaus near -1/2 and -3/2 $e^2/h$ (see supplementary materials). This is common in our devices too, although the parameter windows are always small. Are these really half-integer QH plateaus and a sign of the Landau quantization of a single Dirac channel[2] ? The relevant physics will be discussed in more detail below.

In Fig 1(e) we note that, when the magnetic field is increased to -12 T, the plateau of the device assumes the state n = -1, corresponding to the quantization of both surfaces, $\nu_b$ = -1/2 and $\nu_t$ = -1/2, although the n = 0 state, with $\nu_b$ = 1/2 and $\nu_t$ = -1/2, is well maintained[16, 19]. In Fig 2(b), we note that the QH plateau reaches -1.6 $e^2/h$ at a magnetic field of -6 T.

To understand the physics of the non-integer plateau and their evolution, we measured the full range of data by changing the temperature, gate voltage and magnetic field systematically. RGFD analysis was carried out in the ($G_{xx}$, $G_{xy}$) space for the device before (Fig. S1 in supplementary materials) and after cluster deposition (Fig 2(c)). The CVPs of the RGFD indicate the complete Landau quantization of the bottom surface. It is possible to see two sets of CVPs pointing to $G_{xy}$ of 0 and –$e^2/h$ respectively in this device. Both sets of CVPs clearly show a continuous evolution and finally reach an integer value with increasing field. Note that in all states of the CVPs in RGFD, the bottom surfaces have become quantized. The -3/2 QH plateau is the result of the -1/2 quantized bottom surface and other Hall conductance components incidentally contributing an $e^2/h$, such as a top surface with a semi-QH state[2] as well as some other contributions. The fixed quantized bottom surface and the traveling top surface demonstrate the stepwise quantization achieved in our experiments.

In Fig. 2(d), we plot the two sets of CVPs as curves, which are essentially the quantization trajectories of the top surfaces after subtracting a -1/2 (or 1/2) integer from the QH index of the bottom surfaces[16]. This results in the quantization trajectory of a separate Dirac channel. The quantization of the -1 plateau travels along an anomalous trajectory over a large range of almost one $e^2/h$, which is not seen in the pristine TI devices without Co clusters (Fig. 2(e)). We repeated these measurements in other devices and were able to reproduce the quantization trajectories. Such anomalous quantization trajectories are characteristic of our Co-cluster-decorated TI devices.

**The delayed Landau-level hybridization model**

We now propose a DLLH model to interpret the stepwise LL quantization in the TI devices, where the following two effects are taken into account. First, the Co clusters form some magnetic moments with finite magnetization strengths $M_z(B)$ in a field. The magnetic moments of the Co clusters further polarize the Dirac fermions on the top surface through antiferromagnetic coupling, which results in a sizeable Zeeman gap[18, 30] $m = -JM_z/2\mu_B$, where J is the exchange interaction constant and $\mu_B$ is the Bohr magneton. Second, a medium field enlarges the Zeeman gap and forms the LL quantization. By solving the Dirac equation with a gauge field (see supplementary material for details), we obtain an unexpected zeroth LL with energy $E_0 = -\text{sgn}(B)JM_z/2\mu_B$, which was positioned precisely at the top of the Zeeman gap. Such a magnetization condition is reasonable in our devices[31].

We begin our interpretation by considering the pristine sample in a sufficiently high field, where conventional Landau quantization has been achieved in Dirac fermions (Figs. 3(a, c)). With a decreasing field, the LLs gradually hybridize with each other (Fig. 3(e)) resulting in the smooth evolution of $G_{xy}$ from $-e^2/h$ towards 0, as shown by the red curve in Fig. 2(e). These two effects must be considered following Co cluster deposition. For a sufficiently high field, we can expect no significant difference between the pristine (Fig. 3(c)) and the deposited sample (Fig. 3(d)) because the LL spacing is much larger than the Zeeman gap (Fig. 3(b)). However, for a medium field where the LL spacing reduces to a magnitude comparable with that of the Zeeman gap (Fig. 3(f)), the shifted zeroth LL significantly enlarges the spacing between -1 and 0 LL as a result of the DLLH. Hence, unlike the pristine sample in which the -1 LL hybridizes before it crosses the Fermi energy and $G_{xy}$ undergoes a smooth quantization trajectory towards 0 (Fig. 3(e)), the present -1 LL is well preserved and the $G_{xy}$ of the top surface undergoes a change from -1/2 towards -3/2 $e^2/h$ before its decay towards 0.

This experimental case illustrates the physics of the anomalous quantization trajectory in RGFD (Fig. 2(d)). At around B = -7T, the Hall conductance from the bottom surface is still quantized at -1/2 $e^2/h$ (Fig. 2(e)), while the top surface is non-quantized and exhibits Hall conductance traveling along the anomalous trajectory due to the DLLH effect. Together with other effects such as the lateral cross section, the stepwise quantization leads to the intangible Hall plateaus observed in Fig. 2(a, b, c).

Using DLLH physics, we can extract the width of the LLs (~25.8meV) before the Co cluster deposition, due to impurity scattering, by the critical field (5.6T) in the quantization trajectory (dashed line marked on the red curve in Fig 2(e)). After the deposition of the Co clusters, the QH plateau evolves from -1 to -1.7 $e^2/h$ while the magnetic field decreases from 12 to 6, which shows the width of the LLs at ~31.4 meV at most. The critical field moves to around 6 T due to the delay and the broadened LLs (dashed line marked on the red curve in Fig 2(d)). The Zeeman gap is thus estimated to be nearly 4.7 meV. The separate quantization of the top/bottom surfaces is thus further corroborated by the distinct LL widths. Nevertheless, the greatly disordered top channels still accommodate dissipationless QH transport. This is a priority for the spin-helical Dirac electrons.

## Methods

**Crystal growth and device fabrication.** Well-refined BSTS crystals were grown by melting high-purity elements of Bi, Sb, Te, Se with a molar ratio of 1:1:1:2 at 850 ℃ for 24 hours in evacuated quartz tubes, followed by cooling to room temperature over one week. The bulk carrier density was as low as $4.5\times10^{14}/cm^3$. The crystal was easily cleaved. Following the work process of graphene, the BSTS microflakes were exfoliated and transferred on doped Si substrates coated with 300 nm $SiO_2$. Au electrodes were applied by standard electron-beam lithography and electron beam evaporation. The Co clusters were deposited by a cluster beam system described elsewhere. The thickness of the samples and the clusters were measured using an atomic force microscope.

**Transport measurement and data analysis.** Transport measurements were performed at low temperatures down to 1.8 K with a magnetic field up to 12 T. Standard lock-in amplifiers (Stanford Research System SR830) with a low-frequency (less than 20 Hz) excitation current of 200 nA (Keithley 6221) were used. The high magnetic field experiments were performed at the High Magnetic Field Laboratory, Chinese Academy of Sciences. The sheet resistance was $R_{sh} = R_{xx}(W/L)$, where $W$ and $L$ are respectively the channel width and length between the longitudinal magnetoresistance $R_{xx}$ voltage probes. The conductance was $G_{xx} = R_{sh}/(R_{sh}^2 + R_{xy}^2)$ and $G_{xy} = R_{xy}/(R_{sh}^2 + R_{xy}^2)$. The RGFDs were plotted with $G_{xy}$ as abscissa and $G_{xx}$ as ordinate in ($G_{xy}$, $G_{xx}$) space. $G_{xy}$ and $G_{xx}$ were dependent on the parameters temperature, magnetic field and gate voltage. Every curve in the RGFDs was plotted at a fixed temperature and magnetic field during the scanning of the gate voltage. By changing the temperature (or magnetic field), we obtained the magnetic field (or temperature-) dependent RGFD.

**Acknowledgments**

The authors acknowledge the valuable discussions held with Professor Yong P. Chen at Purdue University and Professors Qianghua Wang, Xiangang Wan at Nanjing University. We would also like to thank the National Key Projects for Basic Research in China (Grant Nos. 2013CB922103), the National Natural Science Foundation of China (Grant Nos. 91421109, 11134005, 11522432,11274003, 11574288, 91622115), the NSF of Jiangsu Province (Grant No. BK20130054) and the Fundamental Research Funds for the Central Universities, for their financial support of this work. We would also like to acknowledge the helpful assistance of the Nanofabrication and Characterization Center at the Physics College of Nanjing University.

31. Here the magnetization of the top surface increases with the field, therefore it is hard to detect it directly and we see some loops. However, we can check its effect by fitting the weak antilocalization using the Hikami-Larkin-Nagaoka formula. We find that the channel parameter a decreases when we increase the fitting range of the magnetic field from (0-0.2T) to (0-1T). Such a decrease can be related to the weak localization contribution from a magnetization gap of ~meV (Rhys. Rev. Lett. 107, 076801 (2011)). This lends support to the validity of the proposed model.

**Figure Captions**

**Figure 1. Surface dominant transport and Quantum Hall effect in the BSTS device**

(**a**) A typical curve showing the temperature-dependent resistance. The inset shows the back-gate measurement configuration and the device's elemental composition. (**b**) atomic force microscopic image of the device. The magnification in (b) shows the granular morphology of the device surface. The sizes of the nanoclusters are tens of nanometers and their height is about 5 nm. The scale bar on the left is 10 μm (1 μm in the zoomed-in version). (**c**) The gate-dependent conductance at different temperatures (2~40K), exhibiting a bipolar characteristic. The minimum conductance appears for a gate voltage of around 16V. (**d**) The sheet resistance of several samples (some of them measured before Co cluster deposition). Red dots show the data measured at 2K, all of which fall within a small range, with or without Co clusters. This indicates the dominance of the surface transport. However, the blue dots measured at 270K fall outside the shaded region. (**e**) The quantum Hall effect observed at a temperature of 1.8 K and a field of 12 T.

**Figure 2. Renormalization group flow diagram (RGFD) analysis with the result of quantization trajectory of a single Dirac channel.**

(**a**) The 3/2 QH plateau observed at -7 T and 1.8 K, obtained while scanning the gate voltage. (**b**) The smaller QH plateau at -6 T and 1.8 K. (**c**) The RGFD analysis in ($G_{xy}$, $G_{xx}$) space based on data measured between -3 and -12 T. The converging points (CVPs), the local minimum with a vanishing $G_{xx}$ and plateau $G_{xy}$, indicate the complete QH filling. Note that the anomalous RGFD trajectory is characteristic of this system. **(d)** We see two sets of CVPs, pointing to $G_{xy}$ of 0 and $-e^2/h$; however, the converging trajectories are obviously different. The inset shows the corresponding plots of $G_{xx}$. The dashed line marks the $G_{xy}$ minimum. **(e)** The CVP trajectories for the sample before the Co cluster deposition. The dashed line marks the $G_{xy}$ minimum.

**Figure 3. The delayed Landau Level (LL) Hybridization (DLLH) model**

**(a, c, e)** and **(b, d, f)** show schematically the LL diagram before/after Co cluster deposition, respectively, during the Hall quantization. The orange dashed lines mark the Fermi energy. Under a magnetic field, the deposited Co clusters induce a Zeeman-like gap in the Dirac cone of the top surface. The zeroth Landau level is shifted to the top of the Zeeman gap (b), while the bottom surface is unchanged, similar to the case for clean devices (a). This makes no significant difference at a sufficiently high field (c, d), while at a medium field, it leads to the observed stepwise quantization (e, f). The anomalous RGFD trajectory (Fig. 2c) is the result of the DLLH of the top surface (f).

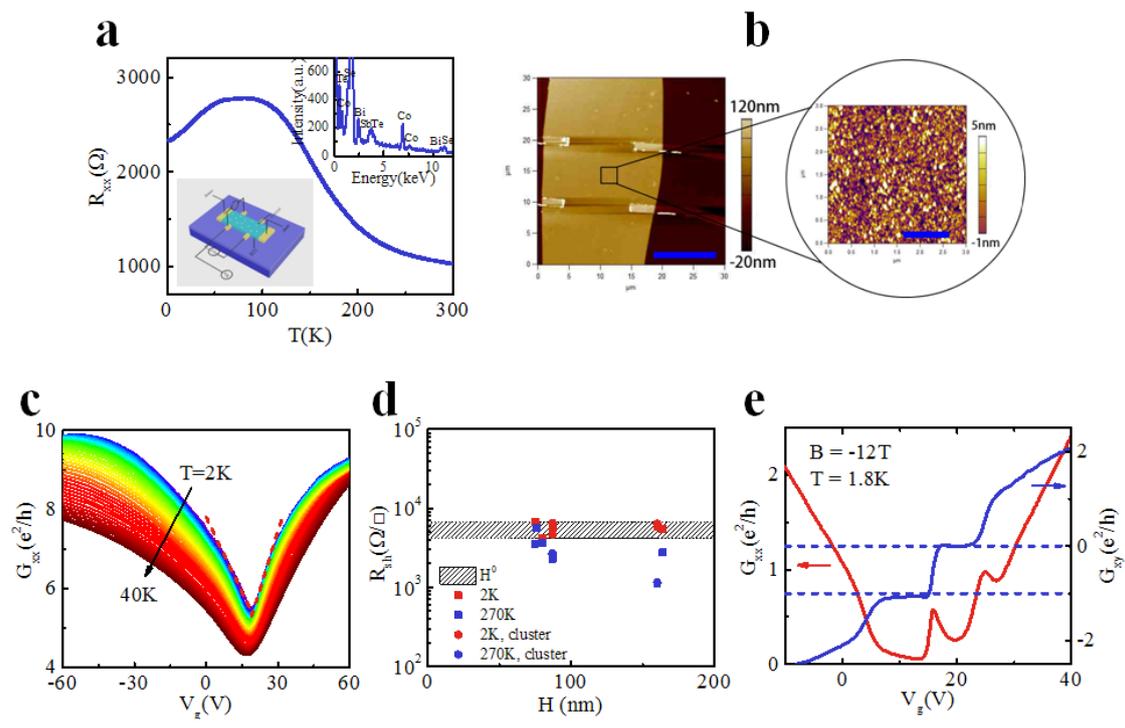

**Zhang et al Figure 1**

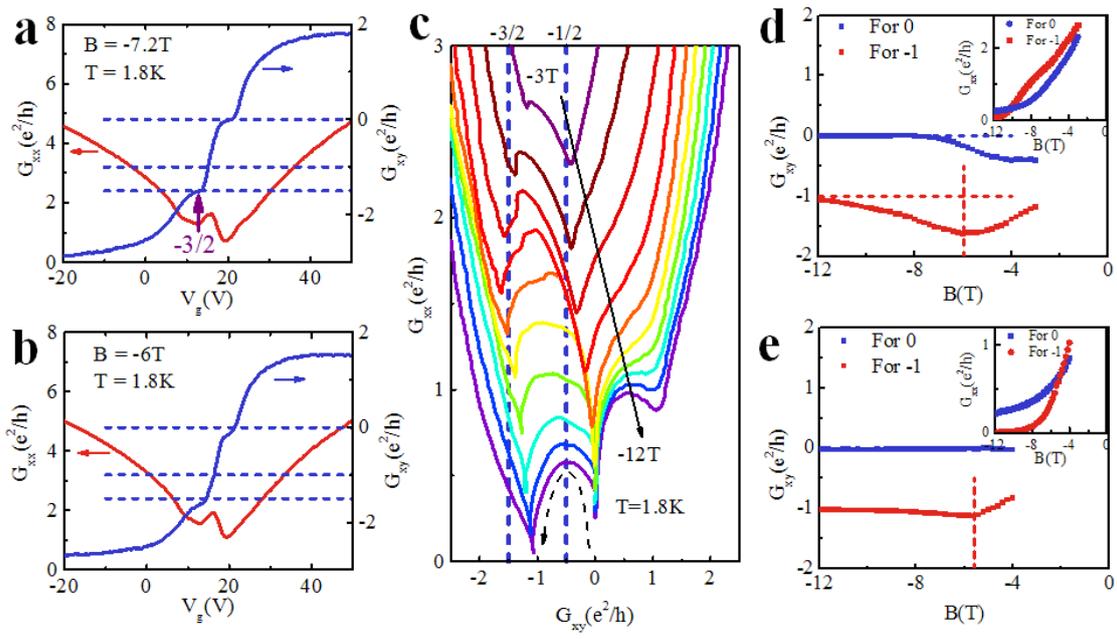

**Zhang et al Figure 2**

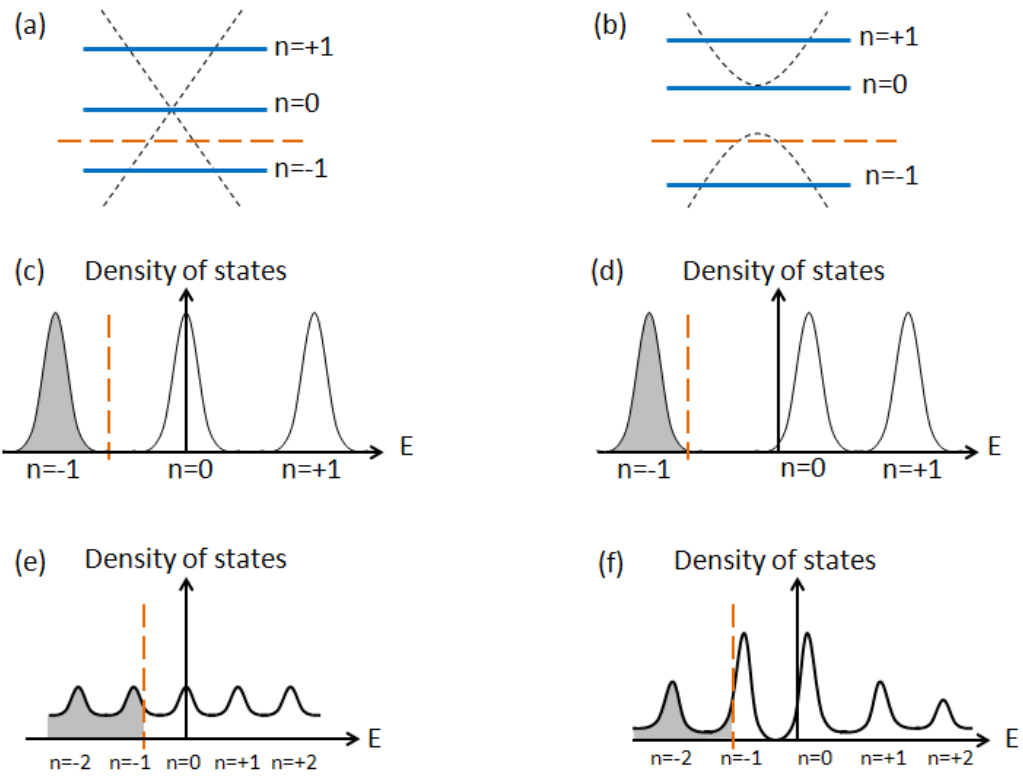

**Zhang et al Figure 3**